\documentclass[12pt,a4paper]{article}

\usepackage{amsmath}
\usepackage{amssymb}
\usepackage{graphicx}
\usepackage{float}

\newcommand{\R}{\ensuremath{\mathbb{R}}}
\newcommand{\bmu}{{\boldsymbol\mu}}
\newcommand{\bsigma}{{\boldsymbol\sigma}}
\newcommand{\rd}{\ensuremath{\mathrm{d}}}
\newcommand{\re}{\ensuremath{\mathrm{e}}}
\newcommand{\E}{\ensuremath{\mathbb{E}}}
\renewcommand{\P}{\ensuremath{\mathbb{P}}}
\newcommand{\C}{\ensuremath{\mathbb{C}}}
\newcommand{\CR}{\ensuremath{\C\setminus\R^+}}
\newcommand{\Nl}{\ensuremath{\mathbb{N}}}
\newcommand{\NC}{\ensuremath{\mathbf{NC}}}
\newcommand{\1}{\ensuremath{\boldsymbol{1}}}
\newcommand{\ord}[1]{\ensuremath{\mathrm{O}\left(#1\right)}}
\newcommand{\Tr}{\mathop\mathrm{Tr}\nolimits}
\newcommand{\exs}[2]{\ensuremath{\E_{#1}\left(#2\right)}}
\newcommand{\h}{{1/2}}
\newcommand{\id}{\ensuremath{\text{1\hspace{-.6ex}I}}}
\font\cmsyfont=cmsy10 at 17pt
\newcommand{\bigtimes}{\mathop{\hbox{\cmsyfont\char2}}}
\newcommand{\probs}[2]{\ensuremath{\P_{#1}\left(#2\right)}}

\newcommand{\cvgwk}{\ensuremath{\stackrel{w}{\rightarrow}}}

\newtheorem{theorem}{Theorem}
\newtheorem{lemma}{Lemma}

\newenvironment{proof}{\setlength{\parindent}{0pt}{\bf
Proof:}\enskip}{\par\hfill $\blacksquare$\par\medskip}
\newenvironment{proofof}[1]{\setlength{\parindent}{0pt}{\bf
Proof of #1:}\enskip}{\par\hfill $\blacksquare$\par\medskip}

\begin{document}

\begin{center}
{\huge\bf The mutual affinity of \\[6pt]
 random measures} 

\vspace{40pt}

{\large M.~Fannes\footnote{Email:
	{\tt mark.fannes@fys.kuleuven.ac.be }} and
P.~Spincemaille\footnote{Email:
	{\tt pascal.spincemaille@fys.kuleuven.ac.be}}
	\footnote{Acknowledges financial support from FWO project G.0239.96}}
\\[7pt]
{Instituut voor Theoretische Fysica}\\
{Katholieke Universiteit Leuven}\\
{Celestijnenlaan 200D}\\
{B-3001 Heverlee, Belgium} 
\end{center}
\vspace{40pt}

\begin{abstract}
We consider a set of probability measures on a finite event space
$\Omega$. The mutual affinity is introduced in terms of the spectrum of
the associated Gram matrix. We show that, for randomly chosen measures,
the empirical eigenvalue distribution of the Gram matrix converges to a
fixed distribution in the limit where the number of measures, together
with the cardinality of $\Omega$, goes to infinity.
\end{abstract}

\newpage

%\keywords{affinity; random measures; random matrices; resolvent method; Marchenko-Pastur distribution} 

% \ams{60G57}{60B12; 62E99} 
%60G57 Probability theory and stochastic processes Stochastic Processes Random
%measures 60B12: Probability theory and stochastic processes Probability theory
%on algebraic and topological structures Limit theorems for vector-valued random
%variables (infinite-dimensional case) 62E99 Statistics Distribution theory None
%of the above, but in this section

\section{Introduction} 

Given two probability measures, there are several ways to define their distance.
This is, \textit{e.g.}, important in problems where a sequence of measures
converges and the nature of this convergence has to be dealt with in a
quantitative way. Common examples are the relative entropy and the total
variation distance. 

Here we shall focus on the \emph{Hellinger distance}. As we shall only consider
finite event spaces, say $\#\Omega=N$ with $N\in\Nl_0$, probability measures
$\bmu$ on $\Omega$ are $N$-tuples $(\mu_1, \mu_2, \ldots, \mu_N)$ which satisfy
$\mu_\alpha \ge 0$ and $\sum_\alpha \mu_\alpha = 1$. The Hellinger distance is: 
\begin{equation*}
 d^2_{\mathrm H}(\bmu_1,\bmu_2) := \frac{1}{2} \sum_{\alpha =1}^N
 (\sqrt{\mu_{1\alpha}} - \sqrt{\mu_{2\alpha}})^2,
\end{equation*}
sometimes, the factor $\frac{1}{2}$ is left out. The Hellinger distance
is a real number between 0 and 1. A related notion is the
\emph{affinity} between two probability measures, defined as
\begin{equation*}
 A(\bmu_1,\bmu_2) := 1 - d^2_{\mathrm H}(\bmu_1,\bmu_2) = \sum_{\alpha=1}^N
 \sqrt{\mu_{1\alpha} \mu_{2\alpha}} = \langle \bmu_1^{1/2}, \bmu_2^{1/2}
 \rangle.
\end{equation*}
In the last term, we have used the short-handed notation $\bmu^{1/2}$ for
the $N$-dimensional vector $(\mu_1^{1/2}, \mu_2^{1/2}, \ldots,
\mu_N^{1/2})$. Two probability measures have affinity one only when
they are equal. Two different degenerate probability measures have
affinity zero. 

Given several measures $\bmu_i$, $i=1,\ldots,K$, one can ask for a 
generalisation of the notion of affinity. The problem is to find a way
of measuring how many of those measures are close to each
other. Here we propose to use the concept of \emph{Gram matrix} 
\begin{equation*}
 G := \left[ A(\bmu_i,\bmu_j) \right]_{i,j=1,\ldots,K}.
\end{equation*}
$G$ is positive semi-definite and its spectrum is independent of the
order of the $\bmu_i$'s. 

A lot of information about the mutual affinities of the probability
measures is encoded in the spectrum of $G$. To appreciate this fact,
let us for a moment consider degenerate probability measures. The
affinity between any two of these can only be one or zero. In the case
all $K$ probability measures are equal, all entries of $G$ are equal to
1 and, therefore, its eigenvalues are $K$ and $0$ with respective 
multiplicities 1 and $K-1$. The other extreme situation is $K$
different degenerate probability measures, in which case $G$ is the
$K$-dimensional identity matrix with eigenvalue 1 occurring $K$ times. 
For an arbitrary set of degenerate measures, \textit{i.e.}\ for a set of 
symbols in $\Omega$, the spectrum of $G$ determines the relative
frequencies of the different symbols appearing in the set.

Of course, allowing general probability measures, any positive number
can be an eigenvalue of $G$, but the general picture remains and can
be described as follows. An eigenvalue distribution which puts a
lot of weight on eigenvalues close to zero indicates that a large group
of probability measures are close to each other (have large mutual
affinities). If, on the other hand, a sizeable portion of the
eigenvalues occur relatively far away from zero, the probability
measures have in general low mutual affinities. 

Here we shall study the Gram matrix for independently and randomly
chosen  probability measures with respect to the uniform distribution
on the simplex $\Lambda_N = \{\bmu = (\mu_1, \ldots, \mu_N) \mid \sum_\alpha
\mu_\alpha =1 \text{ and } \mu_\alpha \geq 0\}$.  The Gram matrix and
its spectrum are now random objects.  We want to study these objects
when both the number of measures and the cardinality of the event space
become large.  More specifically, we study the spectrum of the random 
Gram matrix in the limit $N=\#\Omega \rightarrow \infty$, the number of
measures $K(N)\rightarrow\infty$ and $K(N)/N\rightarrow\tau$ where
$\tau$ is a given positive number. We shall explicitly calculate the
limiting expectation value of the empirical eigenvalue distribution 
\begin{equation*}
 \rho_K(x) := \frac{1}{K}\sum_{i=1}^K \delta(x -\lambda_i),
\end{equation*} 
where $\lambda_1, \ldots,\lambda_K$ are the (random) eigenvalues of the
Gram matrix. We shall, moreover, prove that the convergence occurs with
probability 1.

The setting of this problem is similar to that of the Wishart matrices:
let $A$ be a real random $N\times K$ matrix with $N(0,1)$ i.i.d.\
entries, let $K=\tau N$ for $\tau \geq 0$ and consider the limit
$N\rightarrow\infty$. It is known that the empirical eigenvalue
distribution of the random matrix $A^*A/K $ converges to the
distribution 
\begin{equation}
 \rho_{\mathrm{MP}}(x,\tau) =
 \begin{cases}
   \delta(x-1) & \text{ if } \tau = 0 \\
   \sigma(x,\tau) & \text{ if } 0 < \tau \leq 1 \\
   \displaystyle \frac{\tau-1}{\tau} \delta(x) + \sigma(x,\tau) & \text{ if }
   \tau > 1
 \end{cases}
\label{rhotau}  
\end{equation}
 with
\begin{equation*}
 \sigma(x,\tau) =
 \begin{cases}
   \displaystyle \frac{\sqrt{4\tau x -(x +\tau -1)^2}}{2 \pi \tau x}
   &\qquad (1-\sqrt{\tau})^2 \leq x \leq (1+\sqrt{\tau})^2 \\
   0 &\qquad \text{otherwise.}
 \end{cases}
\end{equation*}
This distribution is known as the \emph{Marchenko-Pastur 
distribution}~\cite{MP} and we shall obtain it in
Theorem~\ref{verw:eigen}. In~\cite{cfp}, the same distribution arose in
the context of Gram matrices associated to random vectors. 

The paper consists of two more parts. In Section~\ref{convexp}, we
discuss some general  features of the spectrum of the random Gram
matrices and calculate the limiting expectation of the empirical
eigenvalue distribution using the Stieltjes transform. The main theorem
in this section is Theorem~\ref{verw:eigen}. In
Section~\ref{almostsure}, we prove that the convergence of the
empirical eigenvalue distribution occurs almost surely. This is the
contents of Theorem~\ref{kans1}.

\section{Convergence in expectation}\label{convexp}

Denote by $\Lambda_N$ the simplex  $\{\bmu = (\mu_1, \ldots,\mu_N)\in \R^N
\mid \mu_\alpha \geq 0 \text{ and }\sum_{\alpha=1}^N \mu_\alpha =
1\}$. It is the space of probability measures on an event space
$\Omega$ with $N$ elements. On this space, a uniform measure $\sigma$
can be put in the sense that 
\begin{equation*}
 \int_{\Lambda_N} f(\bmu)\, \rd\sigma(\bmu) = \frac{1}{|\det(A)|}
 \int_{\Lambda_N} f(A \bmu)\, \rd\sigma(\bmu),
\end{equation*}
for every integrable function $f$ on $\Lambda_N$ supported in
$A\Lambda_N$ and for every invertible stochastic matrix $A$. ($A$ is
stochastic if $A_{\alpha\beta} \geq 0$ and $\sum_\beta A_{\alpha\beta} =
1$). This uniform measure is just the Lebesgue measure on $\Lambda_N$. We
can also obtain this measure in terms of the larger space $(\R^+)^N$ of
which $\Lambda_N$ is a subset. If we choose $N$ independent random
variables $x_i$, all distributed according to the \emph{exponential
distribution} with some fixed mean, then $\bmu :=
(x_1,\ldots,x_N)/(x_1+\cdots+x_N)$ is uniformly distributed on
$\Lambda_N$, a fact which is, \textit{e.g.}, proven in~\cite{sch}.

Now we choose $K$ measures $\bmu_j\in\Lambda_N$, independently and uniformly
distributed, and associate with them the Gram matrix $G$:  
\begin{equation*}
 G = \left[ A( \bmu_i,\bmu_j) \right]_{i,j=1,\ldots K}.
\end{equation*}
We shall study the spectrum of $G$ in the limit $K,N \rightarrow \infty$,
keeping the ratio $K/N =: \tau$ fixed. The Gram matrix is of course a
random object but its spectrum has typical properties. The first
characteristic of the spectrum is the presence of one eigenvalue much
larger than the others. This eigenvalue is the norm of $G$ as $G$  is
positive definite and it grows, as we shall show, linearly with $N$. The
remaining eigenvalues are typically concentrated on an interval close to
zero. In fact we prove:

\begin{theorem}\label{verw:eigen} 
 The empirical eigenvalue distribution $\rho_K(x)$ converges weakly in
 expectation to the Marchenko-Pastur distribution (\ref{rhotau}) scaled with the factor
 $a = 1 - \frac{1}{4}\pi$, i.e.
 \begin{equation*}
  \E\left(\rho_K(x)\right) \cvgwk \frac{1}{a} \rho_{\mathrm{MP}}\Bigl(
  \frac{x}{a} \Bigr)
 \end{equation*}
\end{theorem}

We first prove some lemmas and comment on the (expectation of the) norm
of the random Gram matrix. First, we need expectations of 
arbitrary moments of the components of random probability measures.

\begin{lemma}
 Let $\bmu = (\mu_1,\ldots,\mu_N)$ be a uniformly random probability measure from
 $\Lambda_N$ and let $\alpha_1, \ldots, \alpha_N \geq 0$; then
 \begin{equation}\label{momexp}
  \E\left( \mu_1^{\alpha_1} \cdots \mu_N^{\alpha_N} \right) =
  \frac{(N-1)! \prod_{i=1}^N \Gamma(\alpha_i+1)}{\Gamma(\alpha_1+ \cdots+
  \alpha_N + N)}.
 \end{equation}
\end{lemma}

\begin{proof}
Using the representation of the uniform measure on $\Lambda_N$ in terms
of the exponential distribution with mean 1, we write the expectation as
\begin{equation*}
 \E\left( \mu_1^{\alpha_1} \cdots \mu_N^{\alpha_N} \right) =
 \int_0^\infty \rd x_1 \cdots \int_0^\infty \rd x_N \,
 \re^{-(x_1+\cdots+x_N)} \frac{x_1^{\alpha_1} \cdots
 x_N^{\alpha_N}}{(x_1+\cdots+x_N)^{\alpha_1+\cdots+\alpha_N}}.
\end{equation*}
The change of coordinates
\begin{equation*}
 y_i := \frac{x_i}{x_1+\cdots+x_N}, \quad i=1,\ldots,N-1, \qquad y_N :=
x_N                         
\end{equation*}
transforms the integral into
\begin{align*}
 & \int_0^1 \rd y_1 \int_0^{1-y_1} \rd y_2 \cdots \int_0^{1-y_1-\cdots
 -y_{N-2}} \rd y_{N-1} \int_0^\infty \rd y_N \; y_1^{\alpha_1} \cdots y_{N-1}^{\alpha_{N-1}} y_N^{N-1} \\  
 & \qquad (1-y_1-\cdots -y_{N-1})^{\alpha_N-N}\,  \exp \Bigl( 
 -\frac{y_N}{1-y_1-\cdots-y_{N-1}} \Bigr).  
\end{align*}
Integrating with respect to $y_N$ yields
\begin{equation*}
 \int_0^\infty \rd y_N\, y_N^{N-1}\, \exp \Bigl( 
 -\frac{y_N}{1-y_1-\cdots-y_{N-1}} \Bigr) = (1-y_1-\cdots - y_{N-1})^N
 (N-1)!. 
\end{equation*}
After this step, the successive calculation of the integrals over
$y_{N-1},\ldots,y_1$ can be completed using
\begin{equation*}
 \int_0^x \rd y\, y^p (x-y)^q = x^{1+p+q}\, \mathrm B(p+1,q+1) = x^{1+p+q}
 \frac{\Gamma(p+1)\Gamma(q+1)}{\Gamma(p+q+2)},
\end{equation*}
with $\mathrm B$ the Beta Function. 
\end{proof}

As a first application of this lemma, we compute the expectation of a
single entry in the Gram matrix
\begin{align*}
 \E\left( A(\bmu_1, \bmu_2) \right) 
 & = N \left( \E\left( \sqrt{\mu_{1\alpha}} \right) \right)^2  = N \left( \frac{(N-1)!\; \Gamma(\frac{3}{2})}{\Gamma(N+\frac{1}{2})}
 \right)^2 \\
 & = \frac{\pi}{4} + \frac{1}{16N} + \frac{1}{128 N^2} - \frac{1}{512 N^3}
 + \cdots \\
\end{align*} 
This means that in the $N\rightarrow\infty$ limit, every matrix element
has a non-zero mean and, therefore, the norm of the Gram matrix will grow
linearly with $N$; see \textit{e.g.}~\cite{sil}. It turns out that an
expression for $\E\left( \|G\| \right)$ can be given in terms of the
$R$-transform, a basic notion from free probability; see~\cite{voi,petz}.
To state the result, we need some terminology. 

In non-commutative probability, a random variable is an element
from a unital algebra and expectation values are given by unital linear
functionals $\Phi$ on this algebra. The moments of the random variable
$A$ are $m_n := \Phi(A^n)$ with $n\in \Nl$.  Another sequence of numbers
associated with a random variable are its free cumulants
$(k_n)_{n\in\Nl}$. These are defined in terms of \emph{non-crossing
partitions}. A partition $\pi = \{V_1,\ldots,V_s\}$ of the set
$\{1,\ldots,n\}$ is called \emph{crossing} when there exist numbers $1
\leq p_1 < q_1 < p_2 < q_2 \leq n $ and $1 \leq i<j \leq s$ for which
$p_1,p_2 \in V_i$ and $q_1,q_2 \in V_j$. A partition in which no
crossing occurs is called \emph{non-crossing}. Denote by $\NC(n)$ the
set of all non-crossing partitions on $\{1,\ldots,n\}$.  The free
cumulants are defined  recursively by the equations
\begin{equation*}
 m_n = \sum_{\pi\in\NC(n)} k_\pi,
\end{equation*}
with $k_\pi = k_{\#V_1}\cdots k_{\#V_s}$ for $\pi=\{V_1,\ldots,V_s\}$.
For $n=1,2,3$ the free cumulants are equal to the usual cumulants of
probability theory, where no restriction on the partitions occurs. Only
starting from $k_4$, there is a difference due to the fact that at least
4 different indices are needed to have a crossing. \textit{E.g.,} for a
centred $A$, which means that $\Phi(A)=0$, we find $\Phi(A^4) - 3
\Phi(A^2)^2$ for the usual fourth cumulant, while $k_4 = \Phi(A^4) - 2
\Phi(A^2)^2$.  The relation between the $(k_n)_n$ and $(m_n)_n$ can be
formulated elegantly using formal power series. The first one is the
\emph{Cauchy transform}
\begin{equation*}
 C_A(z) := \sum_{n=0}^{\infty} \frac{m_n}{z^{n+1}} = \Phi((z-A)^{-1})
\end{equation*}
and the second the \emph{$R$-transform}
\begin{equation*}
 R_A(z) := \sum_{n=0}^\infty k_{n+1}z^n.
\end{equation*}
The relation between these two transforms is then given by Voiculescu's
formula
\begin{equation*}
 C_A\left( R_A(z) + \frac{1}{z} \right) = z. 
\end{equation*}

\begin{lemma}\label{lemma2}
 Let $\varphi$ be a normalised vector in a Hilbert space $\mathfrak H$,
 let $X$ be a bounded, linear, self-adjoint operator on $\mathfrak H$
 and let $|\varphi\rangle \langle\varphi|$ denote the operator
 $\psi \mapsto \langle \varphi,\psi \rangle \varphi$.
 The norm of 
 \begin{equation*}
  A(\epsilon) := |\varphi\rangle \langle\varphi| + \epsilon X, \qquad
  \epsilon\in\R
 \end{equation*} 
 is given by the asymptotic series
 \begin{equation*}
  \|A(\epsilon)\| = \sum_{n=0}^\infty k_{n+1}\, \epsilon^n = 1 +
  \epsilon\, R_X(\epsilon),
 \end{equation*}
 \textit{i.e.}, for any $n_0\in\Nl$
 \begin{equation*}
  \|A(\epsilon)\| = \sum_{n=0}^{n_0} k_{n+1}\, \epsilon^n + \mathrm
  o(\epsilon^{n_0}).
 \end{equation*}
 The $(k_n)_n$ are the non-crossing cumulants of $X$ with respect to the
 expectation $\Phi(\cdot) := \langle \varphi,\cdot\varphi \rangle$. 
\end{lemma}

\begin{proof}
For $\epsilon$ sufficiently small, $A(\epsilon)$ will have an
eigenvalue coinciding with its norm. Let $\psi(\epsilon)$ be the
corresponding eigenvector, then 
\begin{equation*}
 (|\varphi\rangle \langle\varphi| + \epsilon X) \psi(\epsilon) =
 \|A(\epsilon)\|\, \psi(\epsilon).
\end{equation*}
The vector $\psi(\epsilon)$ depends continuously on $\epsilon$ and tends
to $\varphi$ when $\epsilon\to 0$. Moreover, $\lim_\epsilon \|A(\epsilon)\| =
1$~\cite{kato}. We can rewrite the eigenvalue equation as 
\begin{equation*}
 \psi(\epsilon) = \langle \varphi,\psi(\epsilon) \rangle\, (\|A(\epsilon)\|
 - \epsilon X)^{-1} \varphi.
\end{equation*}
Multiplying with $\varphi$ and using $\langle \varphi,\psi \rangle\ne 0$
for sufficiently small $\epsilon$ yields
\begin{equation}\label{norm3}
 \langle \varphi, (\|A(\epsilon)\| - \epsilon X)^{-1} \varphi \rangle = 1.
\end{equation}
Then (\ref{norm3}) gives
\begin{equation*}
 C_X\left( R_X(\epsilon) + \frac{1}{\epsilon} \right) = \epsilon = C_X
 \Bigl( \frac{\|A(\epsilon)\|}{\epsilon} \Bigr),
\end{equation*}
which is valid for arbritrary small $\epsilon$ and so
\begin{equation*}
 \|A(\epsilon)\| = 1 + \epsilon R_X(\epsilon).
\end{equation*}
\end{proof}

Using Lemma~\ref{lemma2}, we can compute the asymptotic series for
$\E\left( \|G\|^n \right)$.  Set $\varphi = \1 := \left(
\frac{1}{\sqrt{K}}, \ldots , \frac{1}{\sqrt{K}} \right)$ and $\epsilon =
4/K\pi$; then 
\begin{equation*}
 \epsilon\, G = |\varphi\rangle \langle\varphi| + \epsilon\, X
\quad\text{with}\quad 
 X = G - \frac{1}{\epsilon}\, |\varphi\rangle \langle\varphi|.
\end{equation*}
\textit{E.g.,} for $\E\left( \|G\| \right)$, we get
\begin{equation*}
 \E\left( \|G\| \right) = \frac{K\pi}{4} + \E\left( \langle \1, X \1
 \rangle \right) + \frac{4}{K\pi} \E\left( \langle \1, X^2\1 \rangle -
 \langle \1,X\1 \rangle^2 \right) + \ord{\frac{1}{N}}.
\end{equation*}
Using~(\ref{momexp}) and putting as before $\tau=K/N$
\begin{align*}
 \E\left( \langle \1, X\1 \rangle \right) 
 &= 1 - \frac{\pi}{4} + \tau \frac{\pi}{16} + \ord{\frac{1}{N}}, \\
 \E\left( \langle \1, X^2\1 \rangle \right) 
 &= K \tau \left( \frac{\pi}{4} - \frac{5\pi^2}{64} \right) + \ord{1},
 \quad \mbox{and} \\
 \E\left( \langle \1, X\1 \rangle^2 \right) 
 &= \ord{\frac{1}{N}}.
\end{align*}
We can repeat this procedure to get with arbitrary accuracy the
expectation of any power of the norm of the Gram matrix. 

To study the asymptotic eigenvalue distribution, we could, as a first
step, try to obtain the moments of the limiting distribution as limits of
the expectations of moments of $\rho_K$. The largest eigenvalue of $G$
contributes with a weight $1/K$ in $\rho_K$, but, as this largest
eigenvalue is essentially located around $K\pi/4$, its contribution to
the expectation value of the $n$th moment of $\rho_K$ is of the order
$K^{n-1}$ which leads to a divergence. We must therefore remove that
contribution and study  the expectations of the moments of the
non-normalised distribution 
\begin{equation}\label{nietnorm}
 \rho'_K(x) := \frac{1}{K} \sum_{\lambda_i\neq \|G\|} \delta(x-\lambda_i).
\end{equation}
In the limit, the weight of the largest eigenvalue will become
negligible and we recover a normalised distribution. It turns out
that, in principle, the moments of the limiting distribution can be
obtained by calculating the limit of the expectations of the moments
of~(\ref{nietnorm}) using~(\ref{momexp}). The first two moments yield
\begin{align*}
 m_1^K := \int x \rd\rho'_K(x) 
 &= \E\left( \frac{1}{K} (\Tr G - \|G\|) \right) \\
 &= \frac{1}{K} \left( K - \frac{K\pi}{4} + \ord{1} \right) = 1 -
 \frac{\pi}{4} + \ord{\frac{1}{N}} \\
 m_2^K := \int x^2 \rd\rho'_K(x) 
 &= \E\left( \frac{1}{K} (\Tr G^2 - \|G\|^2) \right) \\
 &= \frac{1}{K} \left( K(\tau+1)(1-\frac{1}{16}) - (\tau+1)(\frac{\pi}{2}
 - \frac{\pi^2}{8} ) + \ord{\frac{1}{N}} \right) \\
 & = (1 + \tau)(1-\frac{\pi}{4})^2  + \ord{\frac{1}{N}}
\end{align*}
These moments coincide of course with those from
Theorem~\ref{verw:eigen}.  The computation is however very hard, as the
$n$th moment of $\rho'_K$ requires $n$ terms in the series expansion of
$\E\left( \|G\| \right)$. A quite complicated combinatorial argument is
already required just to cancel the orders of $N$ larger than one in the
traces of $G$. A much more convenient function of the spectrum of the
Gram matrices is the normalised trace of its resolvent. The following
proof bears some resemblance to the approach presented in~\cite{pas}, but
is technically rather different.

\begin{proofof}{Theorem~\ref{verw:eigen}}
Denote by $\sigma(G)$ the spectrum of $G$ and define for $z \in
\C \setminus \sigma(G)$
\begin{equation*}
 C_K(z) :=  \frac{1}{K} \Tr \frac{1}{G-z} = \int_0^\infty \frac{1}{x-z}
 \rd\rho_K(x).
\end{equation*}
The last equality shows that $C_K$ is the \emph{Stieltjes transform}
of the empirical eigenvalue distribution. 

Let $(e_j)_{j=1,\ldots,K}$ be the standard orthonormal basis of $\C^K$ and
$z\in \C \setminus \sigma(G)$; then
\begin{equation*}
 \frac{1}{K} \Tr \frac{1}{G-z} = \frac{1}{K} \sum_{j=1}^K \langle e_j,
 \frac{1}{G-z}e_j \rangle.
\end{equation*}
Now, for every $j$ in the sum, we peel off the $j$th row and column:
\begin{equation*}
 G = \begin{pmatrix}
      G^{(j)} & \varphi^{(j)} \\
      \langle \varphi^{(j)},\,\cdot\, \rangle & 1 
     \end{pmatrix} \text{ with } \varphi^{(j)} := \left( \langle
 \mu_k^{1/2},\mu_j^{1/2} \rangle \right)_{k\ne j}.
\end{equation*}
This means that we write $\C^K$  as $\C^{K-1} \oplus \C e_j$. The
corresponding form for the resolvent is:
\begin{equation*}
 \frac{1}{G-z}=
 \begin{pmatrix}
  \frac{1}{G^{(j)}-z} +
  \frac{(G^{(j)}-z)^{-1} |\varphi^{(j)}\rangle \langle\varphi^{(j)}|
  (G^{(j)}-z)^{-1}}{1-z - \alpha^{(j)}} 
  &- \frac{(G^{(j)}-z)^{-1} \varphi^{(j)}}{1-z-\alpha^{(j)}} \\*[6pt]
  - \frac{\langle \varphi^{(j)},(G^{(j)}-z)^{-1}\,\cdot\,
  \rangle}{1-z-\alpha^{(j)}} 
  &\frac{1}{1-z-\alpha^{(j)}}
 \end{pmatrix},
\end{equation*}
with 
\begin{equation} 
 \alpha^{(j)} := \langle \varphi^{(j)}, (G^{(j)}-z)^{-1} \varphi^{(j)}
 \rangle.
\label{2}
\end{equation} 
Note that in $\alpha^{(j)}$ the vectors $\varphi^{(j)}$ are the only
place where random variables of the $j$th measure occur. The Stieltjes
transform can then be written as
\begin{equation}
 C_K(z) = \frac{1}{K}\sum_{j=1}^{K} \frac{1}{1-z-\alpha^{(j)}}.
\label{1} 
\end{equation}

We shall now take the limit of the expectation value of~(\ref{1}). 
Therefore, we fix a compact $A \subset \CR$ and $z\in A$.  We first
calculate $\exs{j}{\alpha^{(j)}}$, where the subscript $j$ means that
only the random variables appearing in the $j$th vector will be averaged
out. Let $X = 1/(G^{(j)}-z)$ and use $\E\left( \mu_\alpha^\h\mu_\beta^\h
\right) = \pi/4N$ with $\alpha\ne\beta$ and $\E\left( \mu_\alpha \right)
= 1/N$.
\begin{align*}
 \exs{j}{\alpha^{(j)}}
 &= \exs{j}{\sum_{k,l}^{K} \sum_{\alpha,\beta}^{N} \mu^\h_{j\alpha}
 \mu^\h_{k\alpha} X_{kl}\; \mu^\h_{l\beta} \mu^\h_{j\beta}} \\
 &= \sum_{k,l} \left( \frac{1}{N} \sum_\alpha \mu^\h_{k\alpha} X_{kl}
 \mu_{l\alpha}^\h + \frac{\pi}{4N} \sum_{\alpha\ne\beta} \mu^\h_{k\alpha}
 X_{kl} \lambda_{l\beta}^\h \right) \\
 &= \frac{1}{N} (1-\frac{\pi}{4}) \Tr G^{(j)}X + \frac{\pi}{4N} 
 \langle\gamma^{(j)}, X  \gamma^{(j)}  \rangle,
\end{align*}
with $\gamma^{(j)} := \left( \sum_\alpha \mu^\h_{k\alpha} \right)_{k\ne
j}$.  In Lemma~\ref{lem:vec}, we prove that the expectation (now
averaging over all random variables) of the second term converges to
$\pi/4$, uniformly on $A$. Setting 
\begin{equation*}
 f_K(z) := \E\left( \frac{1}{K} \Tr \frac{1}{G-z} \right) 
 \text{ and } 
 f_K^{(j)}(z) := \E\left( \frac{1}{K} \Tr \frac{1}{G^{(j)}-z} \right),
\end{equation*}
we get that
\begin{align*}
 \E\left( \alpha^{(j)} \right)
 &= \frac{1}{N} \left( 1- \frac{\pi}{4} \right) \Tr \left( \id + z\,
 \frac{1}{G^{(j)}-z} \right) + \frac{\pi}{4} + Q(N,z) \\
 &= \left( 1- \frac{\pi}{4} \right) \left( \tau + z \tau f_K^{(j)} \right) +
 \frac{\pi}{4} + Q(N,z),
\end{align*}
with $Q(N,z)$ converging to zero, uniformly on $A$. In
Lemma~\ref{lem:vec:var}, we show that 
\begin{equation*}
 \exs{j}{{\alpha^{(j)}}^2} = \left( \exs{j}{\alpha^{(j)}} \right)^2 + R(N,z),
\end{equation*}
where $\E\left( R(N,z) \right)$ (averaging over all remaining random variables) converges
to zero, uniformly on $A$.
This allows us to write
\begin{align*}
 \left| \E\left( \frac{1}{1-z-\alpha^{(j)}} \right) - 
 \frac{1}{1-z-\E\left( \alpha^{(j)} \right)} \right| &\leq \E\left( \frac{|\E\left( \alpha^{(j)} \right) - \alpha^{(j)}|}
 {|1-z-\alpha^{(j)}| |1-z-\E\left( \alpha^{(j)} \right)|} \right) \\
 & \leq \frac{1}{|\Im(z)|^2} \underbrace{\sqrt{\E\left( \left( \alpha^{(j)} -
 \E\left( \alpha^{(j)} \right) \right)^2 \right)}}_{\sqrt{\E\left(
 R(N,z) \right)}},
\end{align*}
which goes to zero, uniformly on $A$. We get
\begin{align*}
 f_K(z) = \E\left( C_K(z) \right)
 &= \frac{1}{K} \sum_{j=1}^K \E\left( \frac{1}{1-z-\alpha^{(j)}} \right)= \frac{1}{K} \sum_{j=1}^K \frac{1}{1-z-\E\left( \alpha^{(j)} \right)} + 
 \ord{\frac{1}{N}} \\
 &= \frac{1}{1 - z - \left( 1- \frac{\pi}{4} \right) (\tau + z\tau
 f_K^{(j)}(z)) - \frac{\pi}{4}} + \ord{\frac{1}{N}}.
\end{align*}
Consider for a fixed $z \in A$ the sequence $f_1(z), f_2(z), \ldots$. 
This sequence of complex numbers lies in a compact set, so it must
have a convergent subsequence. Moreover, every convergent subsequence
has the same limit because there is only one number $f(z)$ that satisfies 
both the equation
\begin{equation*}
 f(z) = \frac{1}{a - z - a\tau -az\tau f(z)},
\end{equation*}
with $a = 1- \frac{1}{4}\pi$ and the condition $\Im(z)\Im(f(z)) > 0$. 
From this it is immediately clear that
\begin{equation*}
 \lim_{n\to\infty} f_n(z) = \frac{1}{a} f_{\mathrm{MP}}( \frac{z}{a}),
\end{equation*}
where $f_{\mathrm{MP}}(z)$ is the Stieltjes transform of the Marchenko-Pastur
distribution. Because the convergence in expectation of $f_K(z)$ to
$f(z)$ is uniform on compact subsets of $\CR$, it follows that
$\rho_K(x)$ converges in expectation to $\rho_{\mathrm{MP}}(x)$.
\end{proofof}

In the proof of Theorem~\ref{verw:eigen} we used Lemmas~\ref{lem:vec}
and~\ref{lem:vec:var}. The idea behind their proofs is the following.
Each of the entries in the random Gram matrices has approximately the
same value. The eigenvector belonging to the largest eigenvalue,
\textit{i.e.}\ the norm, of such a matrix, has also nearly constant
entries. Vectors like $\gamma^{(j)}$ defined in Lemma~\ref{lem:vec} are
of this kind. This means that an expression like $\langle \gamma^{(j)},
f(G^{(j)})\gamma^{(j)} \rangle$ is approximately equal to
$f(\|G^{(j)}\|)\, \|\gamma^{(j)}\|^2$. In the sequel, we shall drop the
superscript $(j)$ in $\gamma$ and in $G$ as well, moreover, we shall
replace $K-1$ by $K$ wherever it is not relevant for the result,
\textit{e.g.,} wherever we need quantities estimated up to order 1 in
$K$.  

\begin{lemma}\label{lem:vec}
 Let $\gamma = \left( \sum_{\alpha=1}^N \mu_{k\alpha}^\h
 \right)_{k=1,2,\ldots, K}$; then
 \begin{equation*}
  \lim_{N\rightarrow\infty} \E\left( \frac{1}{N} \langle \gamma,
  \frac{1}{G-z} \gamma \rangle \right) = 1,
 \end{equation*}
 uniformly on compact subsets of $\CR$.
\end{lemma} 

\begin{proof}
We start with the calculation of some useful expectations, all of them just
applications of~(\ref{momexp}). First, 
\begin{equation*}
 \E\left( \|\gamma\|^2 \right) = \frac{\pi\tau}{4} N^2 + \left( 1-
 \frac{\pi}{4} \right) \tau N.
\end{equation*}
Setting 
\begin{equation*}
 \eta := \frac{1}{\sqrt{\E\left( \|\gamma\|^2 \right)}}\, \gamma ,
\end{equation*}
implies $\E\left( \|\eta\|^2 \right) = 1$ and 
\begin{align}
 \E\left( \langle \eta, G\eta \rangle \right) 
 &= \frac{\pi\tau}{4} N + \frac{1}{4} (4- \pi) (1+\tau) +
 \ord{\frac{1}{N}}
\label{verw:norm} \\
 \E\left( \langle \eta, G^2 \eta \rangle \right) 
 &= \frac{\pi^2\tau^2}{16} N^2 + \frac{\pi}{8} (4-\pi) \tau (1+\tau) +
 \ord{1}. 
\label{verw:norm2}
\end{align}
Next, we use the spectral theorem for selfadjoint matrices
\begin{equation*}
 G = \int \lambda \; \rd E(\lambda).
\end{equation*}
Set $\lambda_0 := \E\left( \langle \eta, G\eta \rangle \right)$; the
spectral measure $\rd\|E(\lambda)\eta\|^2$ is, in expectation, very much
concentrated around $\lambda_0$:
\begin{align*}
 \E\left( \int_0^\infty (\lambda - \lambda_0)^2\, \rd\|E(\lambda)\eta\|^2
 \right)
 &= \E\left( \langle \eta, G^2 \eta \rangle \right) - \E\left( \langle
 \eta, G \eta \rangle \right)^2 
 \\
 &=: C = \ord{1}, 
\end{align*}
by~(\ref{verw:norm}) and~(\ref{verw:norm2}). A consequence, using
Tchebyshev's inequality, is
\begin{equation}\label{verw:var:2}
 \int_{|\lambda-\lambda_0|> \lambda_0/2} \rd \|E(\lambda)\eta\|^2 \le
 \frac{4}{\lambda_0^2}\, \E\left( \int_0^\infty (\lambda-\lambda_0)^2\,
 \rd\|E(\lambda)\eta\|^2 \right) = \frac{4C}{\lambda_0^2}.
\end{equation}
Now, we are able to prove the lemma. Consider a compact subset 
$A \subset \CR$ and choose $z\in A$; then
\begin{align}
 \left|	\E\left( \frac{1}{N} \langle \gamma, \frac{1}{G-z}\gamma \rangle
 \right) - 1 \right|
 & \le \left| \E\left( \frac{1}{N} \langle \gamma, \frac{1}{G-z} \gamma
 \rangle \right) - \frac{1}{N} \frac{\E\left(
 \|\gamma\|^2\right)}{\E\left( \langle \eta,G \eta \rangle \right) - z}
 \right| 
\nonumber \\
 & \qquad + \left| \frac{1}{N} \frac{\E\left( \|\gamma\|^2
 \right)}{\E\left( \langle \eta,G\eta \rangle \right) - z}- 1 \right|. 
\label{lem:vec:afsch}
\end{align}
The second term is equal to
\begin{equation*}
 \left| \frac{\frac{1}{4} (4-\pi)(1+\tau) + z}{\frac{1}{4}N \pi\tau - z +
 \ord{1}} \right|,
\end{equation*}
which goes to zero uniformly on $A$. The first term of~(\ref{lem:vec:afsch}) gives
\begin{align}
 & \left| \frac{1}{N} \E\left( \|\gamma\|^2 \right) \E \left(
 \Bigl\langle \eta, \Bigl( \frac{1}{G-z} - \frac{1}{\E\left( \langle
 \eta,G\eta \rangle \right) - z} \Bigr) \eta \Bigr\rangle \right)
 \right| 
\nonumber\\
 & \qquad =  \frac{1}{N} \E\left( \|\gamma\|^2 \right) \left|
 \E\left( \int_0^\infty \left(	\frac{1}{\lambda-z} -
 \frac{1}{\lambda_0-z} \right) \rd\|E(\lambda)\eta\|^2 \right) \right| 
\nonumber \\
 & \qquad \leq \frac{1}{N} \E\left( \|\gamma\|^2 \right) \E\left(
 \int_0^{\lambda_0/2}  \frac{|\lambda_0 - \lambda|}{|\lambda-z|
 |\lambda_0-z|} \rd\|E(\lambda)\eta\|^2 \right)
\nonumber\\
 & \qquad \qquad +  \frac{1}{N} \E\left( \|\gamma\|^2 \right) \E\left( \int_{\lambda_0/2}^\infty 
 \frac{|\lambda_0 - \lambda|}{|\lambda-z| |\lambda_0-z|}
 \rd\|E(\lambda)\eta\|^2 \right).
\label{lem:vec:afsch2}
\end{align}
The first integral is bounded from above by
\begin{equation*}
 \frac{\lambda_0}{|\Im(z)| |\lambda_0-z|} \E\left( \int_0^{\lambda_0/2}
 \rd\|E(\lambda)\eta\|^2 \right)  \leq \frac{4C}{\lambda_0 |\Im(z)|
 |\lambda_0-z|}.
\end{equation*}
The second integral in~(\ref{lem:vec:afsch2}) is, provided
$\Re(z)<\lambda_0/2$, bounded by 
\begin{align*}
 &\frac{1}{|\lambda_0/2 - z|}\, \frac{1}{|\lambda_0-z|}\, 
 \E\left( \int_{\lambda_0/2}^\infty |\lambda_0 -\lambda|
 \rd\|E(\lambda)\eta\|^2 \right) \\
 &\le \frac{1}{|\lambda_0/2 - z|}\, \frac{1}{|\lambda_0-z|}\, 
 \E\left( \left( \int_0^\infty (\lambda_0 - \lambda)^2
 \rd\|E(\lambda)\eta\|^2 \right)^\frac{1}{2} \|\eta\| \right) \\
 &\le \frac{1}{|\lambda_0/2 - z|}\, \frac{1}{|\lambda_0-z|}\, \left(
 \underbrace{\E\left( \int_0^\infty (\lambda_0-\lambda)^2
 \rd\|E(\lambda)\eta\|^2 \right)}_{=C} \underbrace{\E\left(
 \|\eta\|^2\right)}_{=1} \right)^\frac{1}{2} \\
 &= \ord{\frac{1}{N^2}}.
\end{align*}
We conclude that~(\ref{lem:vec:afsch}) can be bounded from above by
\begin{equation*}
 \frac{1}{N} \left( \frac{\pi \tau}{4}N^2 + \left( 1- \frac{\pi}{4}
 \right) \tau N \right) \left( \frac{4C}{\lambda_0|\Im(z)| |\lambda_0 -
 z|} + \frac{\sqrt{C}}{|\lambda_0/2 - z||\lambda_0-z|} \right),
\end{equation*}
which gives a uniform bound on $A$, going to zero when $N\rightarrow\infty$.
\end{proof}

\begin{lemma}\label{lem:vec:var}
With the notations introduced in the proof of Theorem~\ref{verw:eigen}
\begin{equation*}
 \exs{j}{{\alpha^{(j)}}^2} = \left(\exs{j}{\alpha^{(j)}}\right)^2 + R(N,z), 
\end{equation*}
where $\E\left( R(N,z) \right)$ converges to zero, uniformly on compact subsets of $\CR$.
\end{lemma}

\begin{proof}
Using the notation introduced in~(\ref{2}) and still denoting
$1/(G^{(j)}-z)$ by $X$, we compute the expectation with respect to the
random variables appearing in the $j$th random probability measure by
multiple applications of~(\ref{momexp}). We get
{\allowdisplaybreaks
\begin{align*}
 & \exs{j}{\langle \varphi^{(j)}, X \varphi^{(j)} \rangle^2} \\
 & = \exs{j}{\sum_{k,l,m,n} \sum_{\alpha,\beta,\gamma,\delta}
 \mu_{j\alpha}^\h \mu_{k\alpha}^\h X_{kl} \mu_{l\beta}^\h
 \mu_{j\beta}^\h \mu_{j\gamma}^\h \mu_{m\gamma}^\h X_{mn}
 \mu_{n\delta}^\h \mu_{j\delta}^\h} \\
 & = \sum_{k,l,m,n} \left( \sideset{}{'}\sum_{\alpha,\beta,\gamma,\delta}
 \mu_{k\alpha}^\h \right.
 X_{kl} \mu_{l\beta}^\h \mu_{m\gamma}^\h X_{mn} \mu_{n\delta}^\h
 \frac{\pi^2}{16N(N+1)} \\
 & \qquad \qquad + 2 \sideset{}{'}\sum_{\alpha,\beta,\gamma} \mu_{k\alpha}^\h
 X_{kl} \mu_{l\alpha}^\h \mu_{m\beta}^\h X_{mn} \mu_{n\gamma}^\h
 \frac{\pi}{4N(N+1)} \\
 & \qquad \qquad + 4 \sideset{}{'}\sum_{\alpha,\beta,\gamma} \mu_{k\alpha}^\h
 X_{kl} \mu_{l\beta}^\h \mu_{m\gamma}^\h X_{mn} \mu_{n\alpha}^\h
 \frac{\pi}{4N(N+1)} \\
 & \qquad \qquad + \sideset{}{'}\sum_{\alpha,\beta} \mu_{k\alpha}^\h
 X_{kl} \mu_{l\alpha}^\h \mu_{m\beta}^\h X_{mn} \mu_{n\beta}^\h
 \frac{1}{N(N+1)} \\
 & \qquad \qquad + 2 \sideset{}{'}\sum_{\alpha,\beta} \mu_{k\alpha}^\h
 X_{kl} \mu_{l\beta}^\h \mu_{m\beta}^\h X_{mn} \mu_{n\alpha}^\h
 \frac{1}{N(N+1)} \\
 & \qquad \qquad + 4 \sideset{}{'}\sum_{\alpha,\beta} \mu_{k\alpha}^\h
 X_{kl} \mu_{l\alpha}^\h \mu_{m\alpha}^\h X_{mn} \mu_{n\beta}^\h
 \frac{3\pi}{8N(N+1)} \\
 & \qquad \qquad + \left. \sideset{}{'}\sum_{\alpha} \mu_{k\alpha}^\h
 X_{kl} \mu_{l\alpha}^\h \mu_{m\alpha}^\h X_{mn} \mu_{n\alpha}^\h
 \frac{2}{N(N+1)}\right), \\
\end{align*}}
where the symbol $\sideset{}{'}\sum_{\alpha_1,\ldots,\alpha_r}$ means
the sum over all $r$-tuples $(\alpha_1,\ldots,\alpha_r)$ in which no
two entries are equal. Denote the seven restricted sums in this
expression $X'_1, \ldots, X'_7$. Rewriting the expression in terms of
the unrestricted sums, which we shall denote by $X_1, \ldots, X_7$, we
get
\begin{align*}
 &\frac{1}{N(N+1)} \left[ \frac{\pi^2}{4} X_1 + \left( \frac{\pi}{2} -
 \frac{\pi^2}{8} \right) X_2 + \left( \pi - \frac{\pi^2}{4} \right) X_3 +
 \left( 1 - \frac{\pi}{2} - \frac{\pi^2}{16} \right) X_4 \right.\\
 &\qquad + \left. \left( 2 - \pi + \frac{\pi^2}{8} \right) X_5 + \left(
 -\frac{3\pi}{2} + \frac{\pi^2}{2} \right) X_6 + \left( -1 +
 \frac{3\pi}{2} - \frac{3\pi^2}{8} \right) X_7 \right].
\end{align*} 
This can be written as
\begin{align*}
 \frac{1}{N(N+1)}
 &\left[ \left( \left( 1 - \frac{\pi}{4} \right) \Tr G^{(j)}X +
 \frac{\pi}{4} \langle \gamma^{(j)}, X \gamma^{(j)} \rangle \right)^2
 \right. \\
 & + \left( \pi - \frac{\pi^2}{4} \right) \langle
 \gamma^{(j)}, XG^{(j)}X \gamma^{(j)} \rangle + \left( 2 - \pi +
 \frac{\pi^2}{8} \right) \Tr G^{(j)}XG^{(j)}X \\
 & + \left( -\frac{3\pi}{2} + \frac{\pi^2}{2} \right) \sum_{k,l,m,n}
 \sum_{\alpha,\beta} \mu_{k\alpha}^\h X_{kl} \mu_{l\alpha}^\h
 \mu_{m\alpha}^\h X_{mn} \mu_{n\beta}^\h \\
 & + \left. \left( -1 + \frac{3\pi}{2} - \frac{3\pi^2}{8} \right) 
 \sum_{k,l,m,n} \sum_{\alpha} \mu_{k\alpha}^\h X_{kl} \mu_{l\alpha}^\h
 \mu_{m\alpha}^\h X_{mn} \mu_{n\alpha}^\h \right].
\end{align*}
From this, the first statement of the lemma is clear. Now it has to be
proven that the expectation of the remaining terms tends to zero. 

The first term of $R(N,z)$ is 
\begin{align*}
 &\frac{1}{N^2(N+1)} \left( \left( 1 - \frac{\pi}{4} \right) \Tr G^{(j)}X
 + \frac{\pi}{4} \langle \gamma^{(j)}, X \gamma^{(j)} \rangle \right)^2 \\
 &\le \frac{2}{N^2(N+1)} \left( \left( 1 - \frac{\pi}{4} \right)^2 \left(
 \Tr G^{(j)}X \right)^2 + \frac{\pi^2}{16} \langle \gamma^{(j)}, X
 \gamma^{(j)} \rangle^2 \right).
\end{align*}
Now 
\begin{equation*}
 \E\left( \left( \Tr G^{(j)}X \right)^2 \right) \le \frac{1}{|\Im z|^2} (K-1)^2,
\end{equation*}
while also $\E\left( \langle \gamma^{(j)},X\,\gamma^{(j)} \rangle^2 \right)$ is of
order $N^2$. We shall show this using the methods of
Lemma~\ref{lem:vec}. Again  we omit the superscript $(j)$, since
this does not change the result in an essential way. We have
\begin{equation*}
 \E\left( \|\gamma\otimes\gamma\|^2 \right) = \frac{\tau \pi^2}{16} N^4 +
 \frac{1}{2} \left( 1 - \frac{\pi}{4} \right) \tau^2\pi N^3.
\end{equation*}
Set 
\begin{equation*}
 \eta := \frac{1}{\E\left( \|\gamma\otimes\gamma\|^2 \right)^\frac{1}{4}}\, \gamma.
\end{equation*}
This definition ensures that $\E\left( \|\eta\otimes\eta\|^2 \right) =
1$. (Note  that this definition differs slightly from the one given in
the previous lemma. The previous definition would have given $1 +
\ord{N^{-2}}$ for the expectation of the square of the norm of
$\gamma\otimes\gamma$.) Again using~(\ref{momexp}), we have
\begin{align*}
 \E\left( \langle \eta\otimes\eta, G\otimes\id\, \eta\otimes\eta
 \rangle \right) 
 &= \frac{\pi\tau}{4} N + \left( 1 - \frac{\pi}{4}\right) (1+\tau) + 
 \ord{\frac{1}{N}} \\
 \E\left( \langle \eta\otimes\eta, G^2\otimes\id\, \eta\otimes\eta
 \rangle \right) 
 &= \frac{\pi^2\tau^2}{16} N^2 + \frac{\pi\tau}{2} \left( 1 -
 \frac{\pi}{4} \right) (1+\tau)N + \ord{1}. 
\end{align*}
Denote by $E(\lambda_1,\lambda_2)$ the joint spectral family of the commuting
operators $G\otimes\id$ and $\id\otimes G$ and put $\lambda_0 := 
\E\left( \langle \eta\otimes\eta, G\otimes\id\, \eta\otimes\eta
\rangle \right)$.
We then have
\begin{align*}
 \E\left( \int (\lambda_i - \lambda_0)^2 \rd\|E(\lambda_1,\lambda_2)
 \eta\otimes\eta\|^2 \right) 
 &=: C'= \ord{1} \qquad i=1,2 \\
 \E\left( \int_A \rd\|E(\lambda_1,\lambda_2) \eta\otimes\eta\|^2 \right) 
 &\le \frac{8C'}{\lambda_0^2} 
\end{align*}
with $A := \{(\lambda_1,\lambda_2) \mid (\lambda_1 - \lambda_0)^2 +
(\lambda_2 - \lambda_0)^2 \le \lambda_0^2/4\}$. We write 
\begin{align*}
 \E\left( \langle \gamma, X \gamma \rangle^2 \right) &= \E\left( \| \gamma\otimes\gamma \|^2 \right) \E\left( \langle
 \eta\otimes\eta, \frac{1}{G\otimes\id-z}\, \frac{1}{\id\otimes G - z}
 \eta\otimes\eta \rangle \right) \\
 & = \E\left( \| \gamma\otimes\gamma \|^2 \right) \E\left( \int
 \frac{1}{(\lambda_1 - z) (\lambda_2 -z)}\,  \rd\|E(\lambda_1,\lambda_2)
 \eta\otimes\eta\|^2 \right).
\end{align*}
 Then
\begin{align*}
 \left| \E\left( \int_A \frac{1}{(\lambda_1 - z) (\lambda_2 -z)}\, 
 \rd\|E(\lambda_1,\lambda_2) \eta\otimes\eta\|^2 \right) \right|
 &\le \frac{1}{|\lambda_0/2 - z| |\lambda_0/2 - z|},
\end{align*}
and
\begin{align*}
 \left| \E\left( \int_{A^c} \frac{1}{(\lambda_1 - z) (\lambda_2 -z)}\, 
 \rd\|E(\lambda_1,\lambda_2) \eta\otimes\eta\|^2 \right) \right| \le \frac{8C'}{|\Im
 (z)|^2\lambda_0^2}.
\end{align*}
As $\lambda_0$ is of the order $N^2$, these last two inequalities show
that the first term of $R(N,z)$ goes uniformly to zero on compact subsets
of $\CR$.

The second term of $R(N,z)$ contains the matrix element $\langle
\gamma^{(j)}, XG^{(j)}X \gamma^{(j)} \rangle$.  Again in the notation
of Lemma~\ref{lem:vec}, this gives
\begin{align*}
 \left| \E\left( \langle \gamma, XGX \gamma \rangle \right) \right| 
 &= \E\left( \|\gamma\|^2 \right) \left| \E\left( \int_0^\infty
 \frac{\lambda}{(\lambda-z)^2}\, \rd\|E(\lambda) \eta\|^2 \right) \right| \\ 
 &\le \E\left( \|\gamma\|^2 \right) \E\left( \int_0^{\lambda_0/2}
 \frac{|\lambda|}{|\lambda-z|^2}\, \rd\|E(\lambda) \eta\|^2 \right) \\
 &\qquad + \E\left( \|\gamma\|^2 \right) \E\left( \int_{\lambda_0/2}^\infty 
 \frac{|\lambda|}{|\lambda-z|^2}\, \rd\|E(\lambda) \eta\|^2 \right).
\end{align*}
The first term can be bounded from above by
$\E\left( \|\gamma\|^2 \right)\, (\lambda_0 /|\Im(z)|^2)\, (4C/\lambda_0^2)$ by
an application  of formula~(\ref{verw:var:2}). This gives a bound of order
$\ord{N}$. The second term has, provided that $\Re(z) \le \lambda_0/2$, a bound 
\begin{align*}
 & \frac{\E\left( \|\gamma\|^2 \right)}{|\lambda_0/2 - z|^2}\,
 \E\left( \int_{\lambda_0/2}^\infty |\lambda|\, \rd\|E(\lambda)
 \eta\|^2 \right) \\
 &\qquad\qquad \le
 \frac{\E\left( \|\gamma\|^2 \right)}{|\lambda_0/2 - z|^2} \E\left( \left(
  \int_{\lambda_0/2}^\infty \lambda^2\, \rd\|E(\lambda) \eta \|^2
  \right)^\h \|\eta\| \right) \\
 &\qquad\qquad \le
 \frac{\E\left( \|\gamma\|^2 \right)}{|\lambda_0/2 - z|^2}\,
 \left( \E\left( \langle \eta, G^2 \eta \rangle \right) \right)^\h
 \left(\underbrace{\E\left( \|\eta\|^2 \right)}_{=1} \right)^\h,
\end{align*}
which is also, as a consequence of~(\ref{verw:norm})
and~(\ref{verw:norm2}), of order $\ord{N}$. 

The third term of $R(N,z)$ is also of order $\ord{N}$ by
\begin{align*}
\left| \Tr G\frac{1}{G-z} G \frac{1}{G-z} \right|
  &= \left|\Tr \left(\id + \frac{2z}{G-z} + \frac{z^2}{(G-z)^2
}\right)\right| \leq K \left( 1 + \frac{2|z|}{|\Im(z)|} + \frac{|z|^2}{|\Im(z)|^2} \right).
\end{align*}

The fifth term admits the following estimate. Writing $\xi_\alpha$ for the
vector $\left( \mu_{k\alpha}^\h \right)_{k=1,\ldots,K}$, we have 
\begin{align}
 & \left| \E\left( \sum_{k,l,m,n} \sum_{\alpha} \mu_{k\alpha}^\h X_{kl}
 \mu_{l\alpha}^\h \mu_{m\alpha}^\h X_{mn} \mu_{n\alpha}^\h \right) \right|
\nonumber \\
 &\qquad\le \sum_{\alpha} \left| \E\left( \langle \xi_\alpha \otimes
 \xi_\alpha, \frac{1}{G-z}\otimes \frac{1}{G-z} \xi_\alpha \otimes
 \xi_\alpha \rangle \right) \right| 
\nonumber\\
 &\qquad = \sum_\alpha \left| \E\left( \int_0^\infty \int_0^\infty
 \frac{1}{\lambda_1 -z}\, \frac{1}{\lambda_2-z}\,
 \rd\|E(\lambda_1,\lambda_2) \xi_\alpha\otimes\xi_\alpha\|^2 \right)
 \right| 
\nonumber\\
 &\qquad\le N \frac{1}{|\Im(z)|^2} \underbrace{ \E\left( \|\xi_\alpha
 \otimes \xi_\alpha\|^2 \right) }_{= \frac {\tau^2} {1+ 1/N} }.
\label{lem:vec:var:term5}
\end{align}

The fourth term is estimated as
\begin{align*}
 \allowdisplaybreaks
&\left| \E\left( \sum_{k,l,m,n} \sum_{\alpha,\beta} \mu_{k\alpha}^\h
 X_{kl} \mu_{l\alpha}^\h \mu_{m\alpha}^\h X_{mn} \mu_{n\beta}^\h \right)
 \right|\\
 &\leq \E \left( \sum_\alpha \left| \sum_{k,l} \mu_{k\alpha}^\h X_{kl}
 \mu_{l\alpha}^\h \right| \left| \sum_{m,n}\sum_\beta \lambda_{m\alpha}^\h
 X_{mn} \lambda_{n\beta}^\h \right| \right)\\
 &\leq \left[ \E \left( \sum_{k,l,m,n}\sum_\alpha  \mu_{k\alpha}^\h
 X_{kl} \mu_{m\alpha}^\h \overline{X_{m,n}} \mu_{n\alpha}^\h \right)
 \right]^\h \\
 & \qquad \qquad \left[ \E \left( \sum_{k,l,m,n}
 \sum_{\alpha,\beta,\gamma} \mu_{k\alpha}^\h X_{kl} \mu_{l\beta}^\h
 \mu_{m\alpha}^\h \overline{X_{mn}} \mu_{n\gamma}^\h \right) \right]^\h.
\end{align*}
The first factor can be treated like~(\ref{lem:vec:var:term5}), while the second
factor is just $\langle \gamma^{(j)},X G^{(j)} X^\dagger\gamma^{(j)} \rangle$.

Hence, all terms contributing to $R(N,z)$ are of $\ord{N}$ divided by
$N(N+1)$. Therefore, the bound on $R(N,z)$ tends to zero for large dimensions.
\end{proof}

\section{Almost sure convergence}\label{almostsure}

In fact we can prove a stronger result. The empirical eigenvalue
distributions are random measures. The randomness is described by  the
reference probability space  $\bigtimes\limits_{j,\alpha\in\Nl} (\R^+,
\re^{-x} dx)$ through  the realization $\bmu_j =
(x_{j1},\ldots,x_{jN})/(x_{j1} + \cdots + x_{jN})$. We shall denote by
$\P$ expectations with respect to this reference probability
space. 

\begin{theorem}\label{kans1}
 The convergence in Theorem~\ref{verw:eigen} occurs with probability 1.
\end{theorem}

\begin{proof}
We essentially follow the proof in~\cite{haa} for the almost sure
convergence of the empirical eigenvalue distribution of the complex
Wishart matrices, but use a different concentration--of--measure
inequality. We need to show that
\begin{equation*}
 \P\left( \lim_{N,K\rightarrow\infty} \frac{1}{K} \Tr f(G_K) =
 \frac{1}{a} \int_0^\infty f(x)\, \rho_{\mathrm{MP}} \Bigl( \frac{x}{a}
 \Bigr)\, \rd x \right) = 1 
\end{equation*}
with $a = 1- \frac{1}{4}\pi$ and $f$ an arbitrary continuous function on
$\R^+$ vanishing at infinity. We can further restrict ourselves to a
dense subset of such functions, namely, we take for $f$ a differentiable
function on $\R^+$ with compact support. Define the function $g$ by
setting $g(x) := f(x^2)$ for $x \in \R^+$. Then $g$ is also
differentiable with compact support and, like $f$, a Lipschitz function
with constant
\begin{equation*}
 c_1 = \sup_{x\in\R^+} |g'(x)|.
\end{equation*}  
Let $\bmu = \{\bmu_i\}_{i=1}^K$, $\bsigma = \{\bsigma_i\}_{i=1}^K$ denote
two sets of $K$ probability measures in $\Lambda_N$. Define the $N\times
K$ matrix $A_{\bmu}$ by
\begin{equation*}
 A_{\bmu}= \begin{pmatrix}
            \sqrt{\mu_{11}} & \ldots & \sqrt{\mu_{K1}} \\
            \sqrt{\mu_{12}} & \ldots & \sqrt{\mu_{K2}} \\
            \vdots & \ddots & \vdots \\
            \sqrt{\mu_{1N}} & \ldots & \sqrt{\mu_{KN}}
           \end{pmatrix},
\end{equation*}
and analogously for $A_{\bsigma}$. Then $A_{\bmu}^* A_{\bmu}$ is the Gram matrix
associated with the set of measures $\{\bmu_i\}_{i=1}^K$. Define $F : 
\Lambda_N \times \cdots \times \Lambda_N \rightarrow \R$ by
\begin{equation*}
 F(\bmu) := \frac{1}{K} \Tr f(A_{\bmu}^* A_{\bmu}).
\end{equation*}
We want to show that the function $F$ satisfies a Lipschitz condition. Define
$\widetilde{A_{\bmu}}$, $\widetilde{A_{\bsigma}}$ by
\begin{equation*}
 \widetilde{A_{\bmu}} = \begin{pmatrix}
                         0 & A_{\bmu}^* \\
                         A_{\bmu}  & 0
                        \end{pmatrix}
 \text{ and }
 \widetilde{A_{\bsigma}} = \begin{pmatrix}
                            0 & A_{\bsigma}^* \\
                            A_{\bsigma} & 0
                           \end{pmatrix}.
\end{equation*}
Now Lemma~3.5 in~\cite{haa} is used to transport the Lipschitz property
of $g$ on $\R^+$ to $M_{N+K}(\C)_{\mathrm{sa}}$, the set of
$(N+K)$-dimensional complex selfadjoint matrices. This lemma implies
\begin{equation}\label{kans1:lip}
 \|g(\widetilde{A_{\bmu}}) - g(\widetilde{A_{\bsigma}})\|_{\mathrm{HS}} \leq c_1
 \|\widetilde{A_{\bmu}} - \widetilde{A_{\bsigma}}\|_{\mathrm{HS}},
\end{equation}
where $\|A\|_{\mathrm{HS}} := \sqrt{\Tr A^* A}$.
Because 
\begin{equation*}
 \widetilde{A_{\bmu}}^2 = \begin{pmatrix}
                           A_{\bmu}^*A_{\bmu}  & 0 \\
                           0 & A_{\bmu} A_{\bmu}^*
                          \end{pmatrix},
\quad\text{we have}\quad
 g(\widetilde{A_{\bmu}}) = \begin{pmatrix}
                            f(A_{\bmu}^*A_{\bmu}) & 0 \\
                            0 & f(A_{\bmu} A_{\bmu}^*) 
                           \end{pmatrix},
\end{equation*}
and an analogous expression for $g(\widetilde{A_{\bsigma}})$.
Now~(\ref{kans1:lip}) implies
\begin{align*}
 \|f(A_{\bmu}^*A_{\bmu}) - f(A_{\bmu}^*A_{\bsigma})\|_{\mathrm{HS}}^2 
 &+ \|f(A_{\bmu} A_{\bmu}^*) - f(A_{\bmu}A_{\bsigma}^*)\|_{\mathrm{HS}}^2 \\
 &\qquad\le c_1^2\left( \|A_{\bmu} - A_{\bsigma}\|_{\mathrm{HS}}^2 +
 \|A_{\bmu}^* - A_{\bsigma}^*\|_{\mathrm{HS}}^2 \right).
\end{align*}
Since $\|A_{\bmu} - A_{\bsigma}\|_{\mathrm{HS}} = \|A_{\bmu}^* -
A_{\bsigma}^*\|_{\mathrm{HS}}$ and because of the Cauchy-Schwarz
inequality, we have
\begin{equation*}
 |F(\bmu) - F(\bsigma)| 
 \le \frac{1}{\sqrt{K}} \|f(A_{\bmu}^*A_{\bmu}) -  f(A_{\bsigma}^*A_{\bsigma})\|_{\mathrm{HS}} \le c_1 \sqrt{\frac{2}{K}} \|A_{\bmu} - A_{\bsigma}\|_{\mathrm{HS}}. 
\end{equation*}
Now 
\begin{align*}
 \|A_{\bmu} - A_{\bsigma}\|_{\mathrm{HS}}^2 
 &= \sum_{i=1}^K \|\sqrt{\bmu_i} - \sqrt{\bsigma_i}\|^2  = \sum_{i=1}^K
 \sum_{\alpha=1}^N |\sqrt{\mu_{i\alpha}} - \sqrt{\sigma_{i\alpha}}|^2 \le \sum_{i=1}^K \sum_{\alpha=1}^N |\mu_{i\alpha} - \sigma_{i\alpha}|\\
 &  \le \sum_{i=1}^K \sqrt{N} \sqrt{\sum_{\alpha=1}^N
 (\mu_{i\alpha} - \sigma_{i\alpha})^2} = \sqrt{N} \sum_{i=1}^K \|\bmu_i - \bsigma_i\|,
\end{align*}
with the notation $\sqrt{\bmu_i} = (\sqrt{\mu_{i1}},\ldots,\sqrt{\mu_{iN}})$.
Now for arbitrary $t>0$ we have, using this Lipschitz condition,
\begin{align}
 &\P\left( |F(\bmu) - F(\bsigma)|>t \right)= \P\left( |F(\bmu) -
F(\bsigma)|^2>t^2 \right) \leq \P\left( c_1^2 \frac{2}{K} \sqrt{N} \sum_{i=1}^K \|\bmu_i-\bsigma_i\| > t^2 \right) \nonumber \\
 & \qquad = \P\left( \sum_{i=1}^K \|\bmu_i - \bsigma_i\| > \frac{t^2 \tau
 \sqrt{N}}{2 c_1^2} \right) \leq \P\left( \sum_{i=1}^K \|\bmu_i\| + \sum_{i=1}^K
 \|\bsigma_i\|  > \frac{t^2 \tau \sqrt{N}}{2 c_1^2} \right) 
\label{kans1:afsch}
\end{align}
Using Lemmas~\ref{lem:conc} and~\ref{lem:conc:som}, we know that there exist
constants $T\ge 0$ and $c_2 > 0$ such that for $t > 2 T/\sqrt{N}$
\begin{equation*}
 \P\left( \sum_{i=1}^K \|\bmu_i\| > t \right) \le K \exp - \frac{c_2 t N}{2}.
\end{equation*} 
From this, it follows that if we choose $N> 8Tc_1^2/\tau t^2$, the
probability~(\ref{kans1:afsch}) can now be treated analogously as in the proof of
Lemma~\ref{lem:conc:som} yielding:
\begin{align*}
 & \exs{\bmu}{\probs{\bsigma}{\sum_{i=1}^K \|\bsigma_i\| >  \frac {t^2
 \tau \sqrt{N}}{2 c_1^2} - \sum_{i=1}^K \|\bmu_i\|}}  \\
 & \leq K \exp -\Bigl( \frac{c_2}{2} \frac{ t^2 \tau \sqrt{N}}{2 c_1^2}
 N \Bigr) + K^2 \exp - \Bigl( \frac{c_2}{2}\frac{t^2 \tau \sqrt{N}}{2
 c_1^2} N \Bigr) \\
 &\le 2\tau^2 N^2 \exp -\Bigl( \frac{c_2 t^2 \tau N^{3/2}}{4c_1^2} \Bigr).
\end{align*}
Then
\begin{align*}
 \P\left( |F(\bmu) - \E(F)|>t \right)&= \P\left( \exp\left[ \lambda^2 |F(\bmu) - \E(F)|^2 \right] >
 \exp(\lambda^2 t^2 ) \right) \\
 & \leq  \frac{\E\left( \exp\left[ \lambda^2 \left| F(\bmu) -
 \E(F) \right|^2 \right] \right)}{\re^{\lambda^2 t^2}}.
\end{align*}
Take now $N > 32 T c_1^2/\tau t^2$. The function $t\mapsto
\exp\left[ \lambda^2(F(\bmu)-t)^2 \right]$ is convex, so Jensen's inequality implies
\begin{align*}
 & \exs{\bmu}{\exp\left[ \lambda^2 \left| F(\bmu) -
 \exs{\bsigma}{F(\bsigma)} \right|^2 \right]} \leq \exs{\bmu,\bsigma}{\exp\left[ \lambda^2|F(\bmu) -
 F(\bsigma)|^2 \right]}\\
 &\qquad = \int_0^\infty 2 \lambda^2 C \re^{\lambda^2 C^2} \P\left(
 |F(\bmu) - F(\bsigma)|>C \right)\, \rd C \\
 &\qquad \leq \int_0^{t/2} 2 \lambda^2 C \re^{\lambda^2 C^2} \rd C +
 2\tau^2 N^2\int_{t/2}^\infty 2 \lambda^2 C \exp \Bigl( \lambda^2 C^2 -
 \frac{ c_2 C^2 \tau N^{3/2}}{4 c_1^2} \Bigr)\, \rd C.  
\end{align*}
If we choose $\lambda^2 = c_2 \tau N^{3/2}/8 c_1^2$, we get
\begin{align*}
 \P\left( |F(\bmu) - \E(F)|>t \right) \le \frac{2\tau^2 N^2 + \exp
 \frac{c_2 \tau N^{3/2}}{8c_1^2} \frac{t^2}{4}}{\exp \frac{c_2 \tau N^{3/2}}{8 c_1^2}t^2}
 &\le 2 \exp - \frac{3 c_2 \tau N^{3/2} t^2 }{32 c_1^2}.
\end{align*}
An application of the Borel-Cantelli lemma shows that this implies
\begin{equation*}
 \P\left( \lim_{N\rightarrow\infty} |F(\bmu) - \E(F)| \leq t \right) = 1,
\end{equation*}
for arbitrary $t$. This completes the proof.
\end{proof}

\begin{lemma}\label{lem:conc}
 There exist absolute constants $T>0$ and $c>0$ such that for $N$ independent
 exponentially distributed random variables $X_1, \ldots, X_N$ and all $t >
 T/\sqrt{N}$ holds that, with $X = (X_1,\ldots,X_N)$ and $S = \sum_{i=1}^N
 X_i$,  
 \begin{equation*}
  \P\left( \frac{\|X\|}{S} > t \right) \leq \re^{-ctN}.
 \end{equation*} 
\end{lemma}

\begin{proof}
See Theorem~3 and Lemma~1 in~\cite{sch} .
\end{proof}

\begin{lemma}\label{lem:conc:som}
 Suppose that for a random variable $X > 0$ there exist constants $c > 0$ and
 $T \geq 0$ such that for all $t > T$ 
 \begin{equation*}
  \P\left( X > t \right) \leq \re^{-c t};
 \end{equation*}
 then, for $N$ identical independent copies $X_1, \ldots, X_N$ of
 $X$ and for $t > 2T$
 \begin{equation*}
  \P\left( \sum_{i=1}^N X_i > t \right) \leq N \re^{-ct/2}.
 \end{equation*}
\end{lemma}

\begin{proof}
We prove this by induction on $N$. The statement is obviously true for $N=1$
because $t > 2T \geq T $, so $\P\left( X_1>t \right) \le \re^{-ct} \le \re^{-ct/2}$. 
Suppose now that the statement is true for $N-1$ copies, then for $t > 2T$
\begin{align*}
 \P\left( \sum_{i=1}^N X_i > t \right) 
 & = \exs{X_1,\ldots,X_{N-1}}{\probs{X_N}{X_N > t -\sum_{i=1}^{N-1}X_i }} \\
 & = \exs{X_1,\ldots,X_{N-1}}{\probs{X_N}{X_N > t -\sum_{i=1}^{N-1}X_i
 }I_{\{\sum_{i=1}^{N-1} X_i \leq \frac{t}{2}\} } } \\
 & \quad +  \exs{X_1,\ldots,X_{N-1}}{\probs{X_N}{X_N > t -\sum_{i=1}^{N-1}X_i
 }I_{\{\sum_{i=1}^{N-1} X_i > \frac{t}{2}\} }} \\
 & \le \re^{-ct/2} + (N-1) \re^{-ct/2} = N \re^{-ct/2}.
\end{align*}
\end{proof}

\end{document}